\documentclass[superscriptaddress, prl, twocolumn]{revtex4}
\usepackage{amsmath,amssymb,amsthm}
\usepackage{xspace}
\usepackage{graphicx} 
\usepackage{latexsym} 
\usepackage[mathscr]{eucal} 
\usepackage{bm}
\usepackage{dcolumn}
\usepackage{array}
\usepackage[usenames,dvipsnames]{color}
\usepackage{ifpdf}
\usepackage{color}
\usepackage{soul,color}
\soulregister\cite7
\soulregister\ref7
\soulregister\pageref7

\newcounter{tlc}

\newtheorem{theorem}[tlc]{Theorem}

\newtheorem{definition}{Definition}
\newtheorem*{theorem*}{Theorem}
\newtheorem*{lemma*}{Lemma}
\newtheorem*{corollary*}{Corollary}

\DeclareMathAlphabet\mathbfcal{OMS}{cmsy}{b}{n}
\newcommand{\past}[1]{\overleftarrow{#1}}
\newcommand{\future}[1]{\overrightarrow{#1}}
\newcommand{\omni}[1]{\overleftrightarrow{#1}}
\newcommand{\xset}{\mathbfcal{X}}

\newcommand{\bra}[1]{\langle #1|}
\newcommand{\ket}[1]{|#1\rangle}

\newcommand{\ip}[2]{\langle #1 | #2 \rangle}

\newcommand{\et}{$\varepsilon$-transducer }

\newcommand{\jt}[1]{#1}

\begin{document}
\title{Using quantum theory to simplify input-output processes}

\author{Jayne Thompson}
\email{thompson.jayne2@gmail.com}
\affiliation{Centre~for~Quantum~Technologies, National~University~of~Singapore, 3 Science Drive 2, 117543, Singapore}

\author{Andrew~J.~P.~Garner}
\affiliation{Centre~for~Quantum~Technologies, National~University~of~Singapore, 3 Science Drive 2, 117543, Singapore}

\author{Vlatko Vedral}
\affiliation{Atomic~and~Laser~Physics, University~of~Oxford, Clarendon~Laboratory,\\Parks Road, Oxford, OX1 3PU, United Kingdom}
\affiliation{Centre~for~Quantum~Technologies, National~University~of~Singapore, 3 Science Drive 2, 117543, Singapore}
\affiliation{Department of Physics, National University of Singapore, 3 Science Drive 2, Singapore 117543}

\author{Mile Gu}
\email{gumile@ntu.edu.sg}
\affiliation{School of Physical and Mathematical Sciences, Nanyang Technological University, 639673, Singapore}
\affiliation{Complexity Institute, Nanyang Technological University, 639673, Singapore}
\affiliation{Centre~for~Quantum~Technologies, National~University~of~Singapore, 3 Science Drive 2, 117543, Singapore}

\begin{abstract} All natural things process and transform information. They receive environmental information as input, and transform it into appropriate output responses. Much of science is dedicated to building models of such systems -- algorithmic abstractions of their input-output behaviour that allow us to simulate how such systems can behave in the future, conditioned on what has transpired in the past. Here, we show that classical models cannot avoid inefficiency -- storing past information that is unnecessary for correct future simulation. We construct quantum models that mitigate this waste, whenever it is physically possible to do so. This suggests that the complexity of general input-output processes depends fundamentally on what sort of information theory we use to describe them.
\end{abstract}

\maketitle

Every experiment involves applying actions to some system, and recording corresponding output responses. Both inputs and outputs are recorded as classical bits of information, and the system's operational behaviour  can always  be regarded as an input-output process that transforms inputs to outputs. Quantitative science aims to capture such behaviour within mathematical models -- algorithmic abstractions that can simulate future behaviour based on past observations.

There is keen interest in finding the simplest models -- models that replicate a system's future behaviour while storing the least past information \cite{shalizi2001computational,crutchfield1989inferring}. The motivations are two-fold. Firstly from the rationale of Occam's razor,  we should posit no more causes of natural things than are necessary to explain their appearances. Every piece of past information a model requires represents a potential cause of future events, and thus simpler models better isolate the true indicators of future behaviour. The second is practical. As we wish to simulate and engineer systems of increasing complexity, there is always need to find methods that utilize more modest memory requirements.

This motivated systematic methods for constructing such models. The state of the art are $\varepsilon$-transducers, models of input-output processes that are provably optimal -- no other means of modeling a given input-output process can use less past information~\cite{barnett2015computational}. The amount of past information such a transducer requires thus presents a natural measure of the process's intrinsic complexity. This heralded new ways to understand structure in diverse systems, ranging from evolutionary dynamics to action-perception cycles \cite{meddis1988simulation,rieke1999spikes,tishby2011information,gordon2011toward}. Yet $\varepsilon$-transducers are classical, their optimality only proven among classical models.  Recent research indicates that quantum models can more simply simulate stochastic processes that evolve independently of input~\cite{gu2012quantum,mahoney2016occam,tan2014towards}.  Can quantum theory also surpass classical limits in modelling general processes that behave differently on different input?

Here, we present systematic methods to construct quantum transducers -- quantum models that can be simpler than their optimal classical  counterparts. The resulting constructions exhibit significant generality: they improve upon optimal classical models whenever it is physically possible to do so. Our work indicates that classical models waste information unavoidably and this waste can be mitigated via quantum processing.

\section{Framework} We adopt the framework of computational mechanics~\cite{shalizi2001computational,crutchfield1989inferring,barnett2015computational}. \jt{An input-output process describes a system that, at each discrete time-step $t \in \mathbb{Z}$, can be `kicked' in a number of different ways, denoted by some $x^{(t)}$ selected from a set of possible inputs $\mathbfcal{X}$. In response, the system emits some $y^{(t)}$ among a set of possible outputs $\mathbfcal{Y}$. For each possible bi-infinite input sequence $\omni{x} = \ldots x^{(-1)} x^{(0)}x^{(1)}\ldots$, the output of the system can be described by a stochastic process, $\omni{Y} = \dots Y^{(-1)} Y^{(0)} Y^{(1)}\dots$, a bi-infinite string of random variables where each $Y^{(t)}$ governs the output $y^{(t)}$. The black-box behaviour of any input-output process is characterized by a family of stochastic processes, $\{\omni{Y}|\omni{x}\}_{\omni{x} \in \omni{\xset}}$. When the input $\omni{x}$ is governed by some stochastic process $\omni{X}$, the input-output process outputs $\omni{y}$ with probability
\begin{equation}\label{eq:inputoutputprocesses}
P[\omni{Y}\! =\! \omni{y}] =  \sum_{\omni{x}} P[\omni{Y} | \omni{X} \!= \!\omni{x}] P[\omni{X}\! = \!\omni{x}].
\end{equation} Therefore, an input-output process acts as a transducer on stochastic processes -- it takes one stochastic process as input, and transforms it into another~\footnote{We also note that there are some subtleties in defining probability distributions over infinite strings, as in Eq. \eqref{eq:inputoutputprocesses}. These issues have been thoroughly discussed in complexity theory literature, for more details see \cite{barnett2015computational}, \cite{kallenberg2006foundations}.}}.

\jt{Computational mechanics typically assumes processes are causal and stationary. Causality implies the future inputs do not retroactively affect past outputs. That is, for all $L \in \mathbb{Z}^+$, we require $P(Y^{(t:t+L)} |\omni{X}) = P(Y^{(t:t+L)} | \past{X}^{(t+L+1)})$, where $\past{X}^{(t+L+1)} = \dots  X^{(t+L-1)}X^{(t+L)}$ and $Y^{(t:t+L)} = Y^{(t)}  \ldots Y^{(t+L)}$.  This naturally bipartitions each stochastic process $\omni{Y}$ into two halves, $\past{y}^{(t)} =  \ldots y^{(t-2)}y^{(t-1)}$ to represent events in the past of $t$, and $\future{y}^{(t)} = y^{(t)} y^{(t+1)}\ldots$ to describe events in the future, governed respectively by $\past{Y}^{(t)}$ and $\future{Y}^{(t)}$.
Stationarity implies the process is invariant with respect to time translation, such that $P(Y^{(t:t+L)}| \omni{x}^{(t)}) = P(Y^{(0:L)} | \omni{x}^{(0)})$ and $P(\omni{Y}^{(t)} | \omni{x}^{(t)} ) = P(\omni{Y}^{(0)} | \omni{x}^{(0)})$, where $\omni{x}^{(t)} = \dots x^{(t-1)} x^{(t)} x^{(t+1)}\dots $ is governed by $\omni{X}^{(t)}$. Hence we can take the present to be $t = 0$, and omit the superscript $(t)$.}

\begin{figure}
\includegraphics[width=0.5\textwidth]{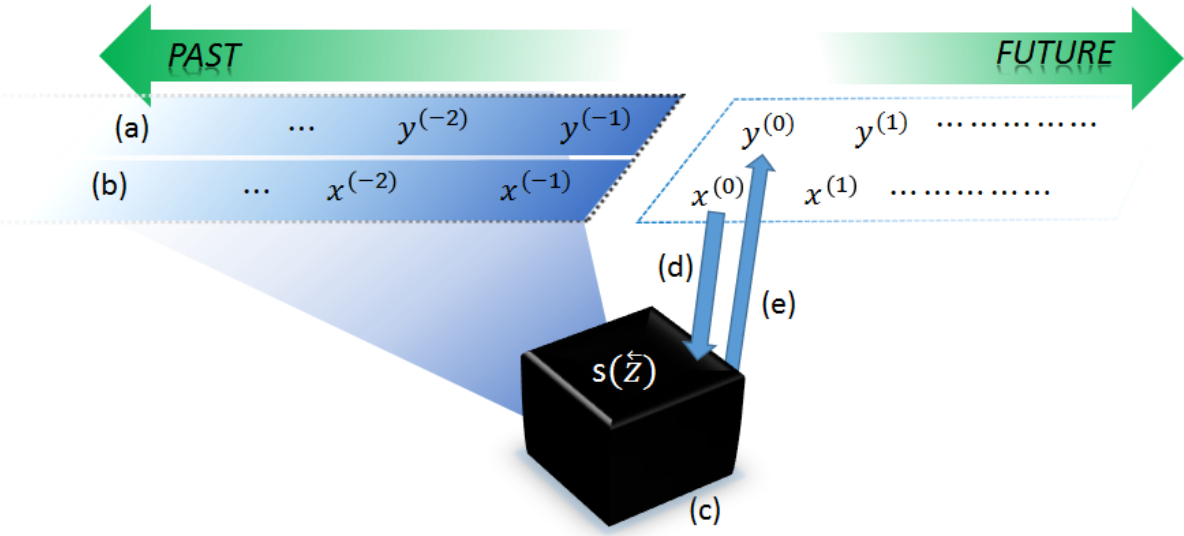}
\caption{\label{fig:modelling} \textbf{Modelling a general input-output process}. Each instance of an input-output process features some specific sequence of past inputs (a) and past outputs (b). A model of such a process describes a systematic method of storing relevant information within a physical system (c), such that for any future input (d), it can replicate the correct statistical output (e).}
\end{figure}

\jt{Each instance of an input-output process has some specific past $\past{z} = (\past{x},\past{y})$.  On future input $\future{x}$, the process will then exhibit a corresponding conditional future governed by $P[\future{Y} | \future{X} \! = \! \future{x}, \past{Z} \! = \! \past{z}]$. A mathematical model of the process should replicate its future black-box behaviour when given information about the past.  That is, each model records $s(\past{z})$ in some physical memory $\Xi$ in place of $\past{z}$, such that upon future input $\future{x}$ the model can generate a random variable $\future{Y}$ according to $P[\future{Y} | \future{x}, \past{z}]$ (see Fig. \ref{fig:modelling}).}

\textbf{Simplest Classical Models.} \jt{Numerous mathematical models exist for each input-output process. A brute force approach involves storing all past inputs and outputs. This is clearly inefficient. Consider the trivial input-output process that outputs a completely random sequence regardless of input. Storing all past information would take an unbounded amount of memory. Yet this process can be simulated by flipping an unbiased coin - requiring no information about the past.}

A more refined approach reasons that replicating future behaviour does not require differentiation of pasts with statistically identical future behaviour. Formally, we define the equivalence relation $\past{z} \sim_{\varepsilon} \past{z}'$ whenever two pasts, $\past{z}$ and $\past{z}'$, exhibit statistically coinciding future input-output behaviour, i.e., whenever $P[\future{Y} | \future{X},\past{Z}\!  =\!  \past{z}] = P[\future{Y} | \future{X}, \past{Z}\!  = \! \past{z}']$. This partitions the space of all pasts into  equivalence classes $\mathbfcal{S} = \{s_i\}$. Each $s_i \in \mathbfcal{S}$ is known as a causal state, and $\varepsilon$ denotes the encoding function that maps each past to its corresponding causal state. \jt{In general, a process can have an infinite number of causal states. Classical studies generally concentrate on cases where $n = |\mathbfcal{S}|$ is finite. In light of this, we have focused our presentation on such cases.}

This motivates the $\varepsilon$-transducer, which stores the causal state $\varepsilon(\past{z})$ in place of $\past{z}$. It then operates according to the transition elements
\begin{equation}
T_{ij}^{y|x} = P[S^{(t)}= s_j, Y^{(t)} \! = \! y | S^{(t-1)}\!  =\!  s_i, X^{(t)}\!  =\!  x];
\end{equation}
the probability a transducer in causal state $s_i \in \mathbfcal{S}$ will transition to causal state $s_j \in \mathbfcal{S}$ while emitting $y \in \mathbfcal{Y}$, conditioned on receiving input $x \in \mathbfcal{X}$. Note that this construction is naturally \emph{unifiliar} --  given the state of the transducer at the current time-step, its state at the subsequent time-step can be completely deduced by observation of the next input-output pair~\cite{barnett2015computational}. Thus iterating through this procedure generates output behaviour statistically identical to that of the original input-output process.

For each stationary input process $\omni{X}$, causal state $s_i$ occurs with probability $p_X\!(i) =P[\varepsilon{(\past{z})} = s_i] $. The $\varepsilon$-transducer will thus exhibit internal entropy
\begin{equation}\label{eq:classicalstatcomplexity}
C_X = -\sum_i p_X\!(i) \log   p_X\!(i).
\end{equation}
$\varepsilon$-transducers are the provably simplest classical models -- any other encoding function $s(\past{z})$ that generates correct future statistics, will exhibit greater entropy~\cite{barnett2015computational}.

\jt{Complexity theorists regard $C_X$ as a quantifier of complexity \cite{barnett2015computational}, the rationale being that it characterizes the minimum memory any model must store when simulating $\{\omni{Y}|\omni{x}\}_{\omni{x} \in \omni{\mathbfcal{X}}}$ on input $\omni{X}$. More precisely, consider the simulation of $N$ such input-output processes, where each instance is driven by $\omni{X}$. $C_X$ then specifies that in the asymptotic limit ($N \rightarrow \infty$) we can use  the $\varepsilon$-transducer to replicate the future statistics of the ensemble by storing the past within a system of $NC_X$ bits.}

In the special case where the input-output process is input independent (i.e., $\future{Y}|\omni{x}$ is the same for all $\omni{x}$), the causal states are reduced to equivalences classes on the set of past outputs $\past{\mathbfcal{Y}}$. $T^{y|x}_{ij}$ become independent of $x$ and is denoted $T^{y}_{ij}$. Here, the $\varepsilon$-transducers are known as $\varepsilon$-machines and $C_X$ as the \emph{statistical complexity}. This measure has been applied extensively to quantify the structure of various stochastic processes~\cite{tivno1999extracting, larrondo2005intensive,gonccalves1998inferring,park2007complexity}.

For general input-output processes, $C_X$ is $\omni{X}$-dependent and is known as the \emph{input-dependent statistical complexity}. In certain pathological cases (e.g.\ always inputting the same input at every time-step), the transducer may have zero probability of being in a particular causal state, potentially leading to a drastic reduction in $C_{X}$. Here, we consider \emph{non-pathological} inputs, such that the transducer has non-zero probability of being in each causal state, i.e., $p_X\!(i) > 0$ for all $i$. It is also often useful to quantify the intrinsic structure of input-output processes without referring to a specific input process~\cite{barnett2015computational}. One proposal is the  \emph{structural complexity},  $\overline{C}= \sup_{X} C_X$, which measures how much memory is required in the worst case scenario.

\jt{\textbf{Example.} We illustrate a simple example of an \emph{actively perturbed coin}. Consider a box with two buttons, containing a single coin. At each time-step, the box accepts a single bit $x \in \{0,1\}$ as input representing which of the two buttons is pressed. In response, it flips the coin with probability $p$ if $x = 1$, and probability $q$ if $x = 0$, where $0 < p,q < 1$. The box then outputs the new state of the coin, $y$. The behaviour of the box is described by an input-output process.}

\begin{figure}
\includegraphics[width=0.5\textwidth]{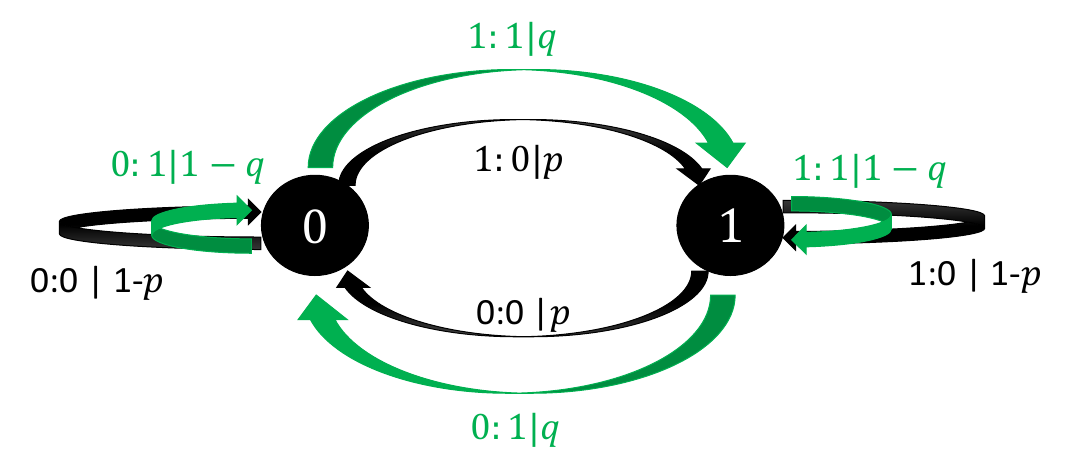}
\caption{\label{fig:causalstatediagram} \jt{\textbf{$\epsilon$-transducer for the actively perturbed coin}. Here, node $0$ ($1$) is identified with causal state $s_0$ ($s_1)$ and represents pasts where the last coin toss resulted in tails (heads). Each non-zero transition element $T_{kj}^{y|x}$ is then represented by a directed edge from node $k$ to $j$ labeled by `$y|x:T_{kj}^{y|x}$'. Here, future statistics can be generated by transition elements $T^{1|0}_{01} = T^{0|0}_{10} = p$, $T^{0|0}_{00} = T^{1|0}_{11} = 1-p$, $T^{0|1}_{10} = T^{1|1}_{01} = q$, $T^{0|1}_{00} = T^{1|1}_{11} = 1-q$.}}
\end{figure}

\jt{First note that when $p = q = 0.5$, the output of the device becomes completely random, all pasts collapse to a single causal state, and the statistical complexity is trivially zero. In all other cases, the past partitions into two causal states, $s_k = \{\past{z}: y^{(-1)} = k\}$, $k = 0,1$, corresponding to the two possible outcomes of the previous coin toss. Future statistics can then be generated via appropriate transition elements. For details, see the graphical representation in Fig. \ref{fig:causalstatediagram}.}

\jt{Consider a simple input process $\omni{X}_u$ where $x = 1$ is chosen with some fixed probability $u$ at each time-step, the symmetry of the transition elements implies  $s_0$ and $s_1$ occur with equiprobability. Thus $C_{X_u} = 1$. Furthermore, as this is the maximum entropy a two-state machine can take, the classical structural complexity of the actively perturbed coin, $\overline{C}$, is also $1$.}

\textbf{Classical Inefficiency.}\jt{ Even though the $\epsilon$-transducer is classically optimal, it may still store unnecessary information. Consider the example above. The $\epsilon$-transducer must store the last state of the coin (i.e., whether the past is in $s_0$ or $s_1$). However, irrespective of input $x$, both $s_0$ and $s_1$ have nonzero probability of transitioning to the same $s_k$ while emitting the same output $y$. Once this happens, some of the information being used to perfectly discriminate between $s_0$ and $s_1$ will be irrevocably lost -- i.e. there exists no systematic method to perfectly retrodict whether an \et was initialized in $s_0$ or $s_1$ from its future behaviour, regardless of how we choose future inputs. Thus the transducer appears to store information that will never be reflected in future observations, and is therefore wasted.}

\jt{We generalize this observation by introducing {\it step-wise inefficiency}. Consider an $\varepsilon$-transducer equipped with causal states $\mathbfcal{S}$ and transition elements $T_{ij}^{y|x}$. Suppose there exists $s_i, s_j \in \mathbfcal{S}$ such that irrespective of input $x$, both $s_i$ and $s_j$ have non-zero probability of transitioning to some coinciding $s_k \in \mathbfcal{S}$ while emitting a coinciding output $y \in \mathbfcal{Y}$ -- i.e., for all $x \in \mathbfcal{X}$  there exists $y \in \mathbfcal{Y}, s_k \in \mathbfcal{S}$ such that both $T_{ik}^{y|x}$ and $T_{jk}^{y|x}$ are non-zero. This implies that at the subsequent time-step, it is impossible to infer which of the two causal states the transducer was previously in with certainty. Thus some of the information being stored during the previous time-step is inevitably lost. We refer to any $\varepsilon$-transducer that exhibits this condition as being \emph{step-wise inefficient}}.

\section{Results} Quantum processing can mitigate this inefficiency. Whenever the $\varepsilon$-transducer of a given input-output process is step-wise inefficient, we can construct a quantum transducer that is provably simpler. Our construction assigns each $s_i$ a corresponding quantum causal state
\begin{equation}\label{eq:quantumcausalstates}
\ket{s_i} = \bigotimes_x \ket{s_i^x}, \,\, \textrm{with}\,\, \ket{s_i^x} = \sum_k \sum_y \sqrt{T_{ik}^{y|x}} \ket{y} \ket{k},
\end{equation}
where $\bigotimes_x$ represents the direct product over all possible inputs $x\in \mathbfcal{X}$, while $\ket{y}$ and $\ket{k}$ denote some orthonormal basis for spaces of dimension $|\mathbfcal{Y}|$ and $|\mathbfcal{S}|$ respectively. The set of quantum causal states $\{\ket{s_i}\}_i$ form a state space for the quantum transducer. Note that although each state is represented as a vector of dimension $(|\mathbfcal{Y}||\mathbfcal{S}|)^{|\mathbfcal{X}|}$, in practice they span a Hilbert space of dimension at most $n = |\mathbfcal{S}|$, \jt{ whenever $|\mathbfcal{S}|$ is finite}. In such scenarios, the quantum causal states can always be stored losslessly within a $n$-dimensional quantum system (see subsequent example and methods).

For an input process $\omni{X}$, the quantum transducer thus has input-dependent complexity of
\begin{equation}
Q_X = -\textrm{Tr}\left[\rho_X\log{\rho_X}\right],
\end{equation}
where $\rho_X = \sum_i p_{X}\!(i) \, \ket{s_i}\bra{s_i}$. \jt{In general $Q_X \le C_X$ \cite{nielsen2010quantum}}. Note that when there is only one possible input, \jt{$\ket{s_i} = \ket{s_i^{0}}$}. This recovers existing quantum $\varepsilon$-machines that can model autonomously evolving stochastic processes better than their simplest classical counterparts~\cite{gu2012quantum}. Our transducers generalize this result to input-output processes. We show that they field the following properties:

\begin{enumerate}
\item \textit{Correctness:} For any past $\past{z}$ in causal state $s_i$, a quantum transducer initialized in state $\ket{s_i}$ can exhibit correct future statistical behaviour. i.e., there exists a systematic protocol that, when  given $\ket{s_i}$ generates $\future{y}$ according to $P[\future{Y}|\future{X} = \future{x},  \past{Z} =\past{z}]$, for \jt{each possible sequence} of future inputs $\future{x}$.
     \item \textit{Reduced Complexity:} $Q_X  < C_X$ for all non-pathologic input processes $\omni{X}$, whenever the process has a step-wise inefficient $\varepsilon$-transducer.
\item \textit{Generality:} Quantum transducers store less memory whenever it is physically possible to do so. Given an input-output process, either $Q_X < C_X$ for all non-pathological $\omni{X}$, or there exists no physically realizable model that does this.
\end{enumerate}

\begin{figure}
\includegraphics[width=0.5\textwidth]{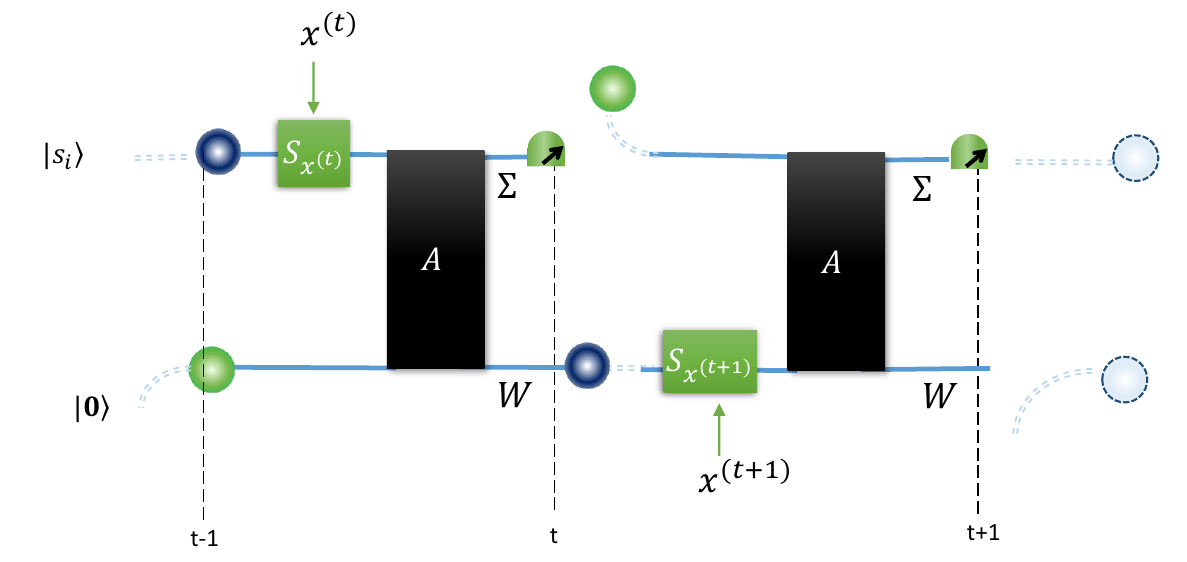}
\caption{\label{fig:quantumsimulation}\jt{ The quantum circuit which illustrates how a quantum transducer initialized in \jt{$\ket{s_i}$ at time $t-1$}, simulates the future behaviour, when supplied with an input sequence $x^{(t)} x^{(t+1)}\dots$. Upon receiving $x$ at time-step $t$, it applies a selection operator $S_{x}: \ket{s_i} \rightarrow \ket{s_i^{x}}$, followed by the quantum operation $A$ that takes each $\ket{s_i^x}\bra{s_i^x}$ to some bipartite state $\sum_{y,k} T^{y|x}_{ik}  \ket{y} \ket{s_k}\bra{y}\bra{s_k}$ with bipartitions $\Sigma$, spanned by $\ket{y}$, and $\mathcal{W}$, spanned by $\ket{s_k}$ (This is always possible with suitable ancilla, see methods). $\Sigma$ is emitted as output, while $\mathcal{W}$ is retained as the quantum causal state at the next time-step. Measurement of $\Sigma$ in the $\{\ket{y}\}$ basis by any outside observer yields outcome $y^{(t)}$. Iterating this procedure then generates correct outputs at each future time-step.}}
\end{figure}

\emph{Correctness} guarantees quantum transducers behave statistically identically to their classical counterparts -- and are thus operationally indistinguishable from the input-output processes they simulate. The proof is done by explicit construction (see Fig.\ \ref{fig:quantumsimulation} and details in the methods).

\emph{Reduced complexity} implies step-wise inefficiency is sufficient for quantum transducers to be simpler than their classical counterparts. The proof involves showing that if for any potential input $x$, both $s_i$ and $s_j$ have non-zero probability of transitioning to some coinciding causal state $s_k$ while emitting identical $y$, then $\ket{s_i}$ and $\ket{s_j}$ are non-orthogonal (see methods). Thus provided two such causal states exist (guaranteed by step-wise inefficiency of the transducer), and each occur with some non-zero probability (guaranteed by non-pathology of the input), $Q_X  < C_X$.

\emph{Generality} establishes that step-wise inefficiency is a necessary condition for any physically realizable quantum model to outperform its classical counterpart. Combined with `reduced complexity', they imply that step-wise inefficiency is the sole source of avoidable classical inefficiency and that our particular quantum transducers are general in mitigating this inefficiency. The proof is detailed in the methods, and involves showing that any model which improves upon an $\varepsilon$-transducer that is step-wise efficient allows perfect discrimination of non-orthogonal quantum states.

Together, these results isolate step-wise inefficiency as the necessary and sufficient condition for quantum models to be simpler than their classical counterparts, and furthermore, establish an explicit construction of such a model. It follows that whenever the $\varepsilon$-transducer is step-wise inefficient, the upper-bound,
\begin{equation}
\overline{Q} = \sup_{X} Q_{X},
\end{equation}
will be strictly less than $\overline{C}= \sup_{X} C_{X}$, provided $\sup_{X} C_{X}$ is attained for a non-pathological $\omni{X}$. Intuitively, this clause appears natural. If an agent wished to drive a transducer to exhibit the greatest entropy, then it would be generally advantageous to ensure the transducer has finite probability of being in each causal state. Nevertheless, as the maximization is highly non-trivial to evaluate, this remains an open conjecture.

\begin{figure}
\includegraphics[width=0.5\textwidth]{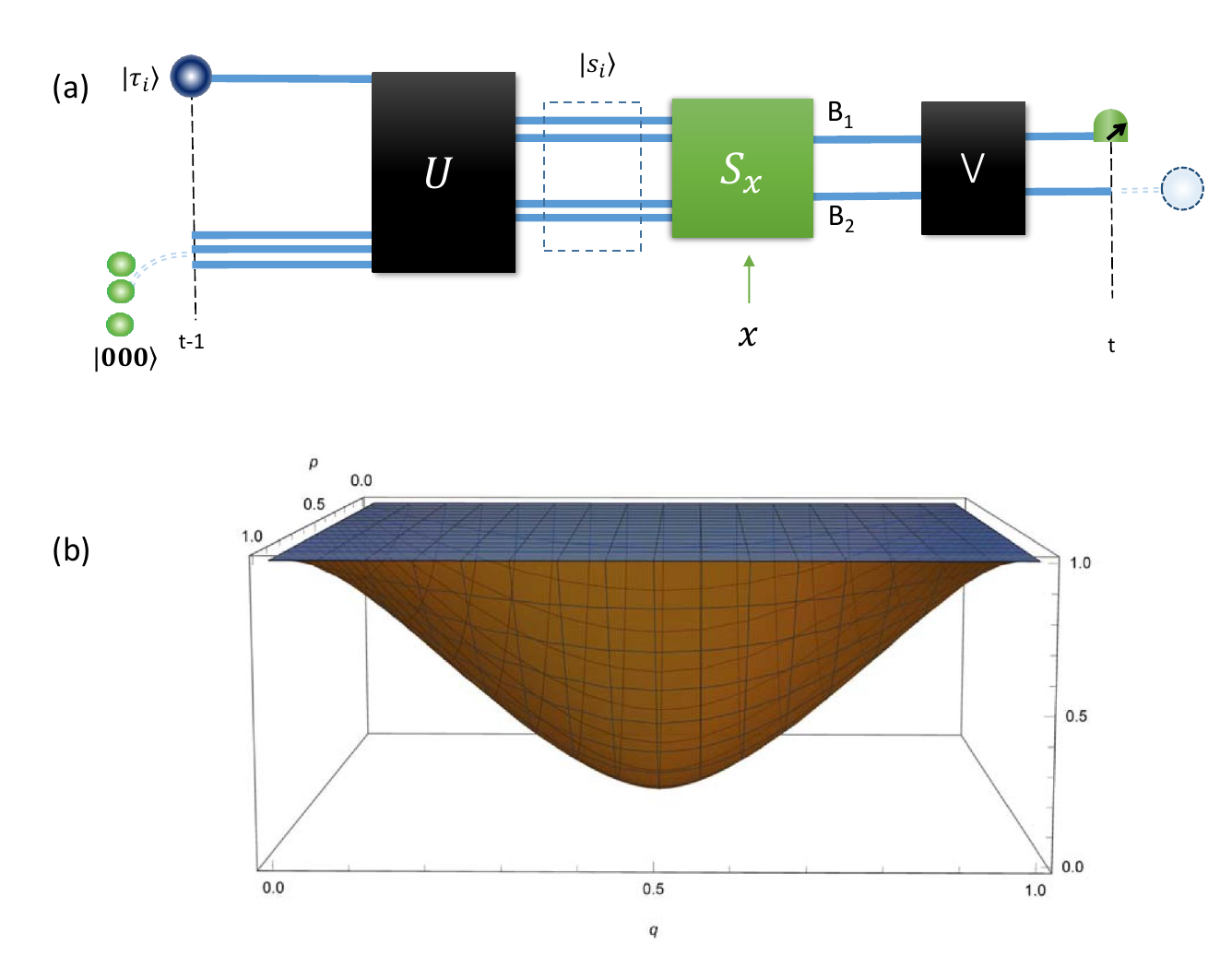}
\caption{\label{fig:coinflip} \jt{The quantum transducer for the perturbed coin can generate appropriate future statistics via the quantum circuit in (a). Suppose a transducer, in state $\ket{\tau_i}$, receives input $x$ at time $t$. To generate output $y$, it transforms $\ket{\tau_i}$ to a 4-qubit quantum state $\ket{s_i}$ by application of an appropriate unitary $U$ on $\ket{\tau_i}$ and three ancilla qubits in state $\ket{000}$. The transducer then discards qubits one and two if $x =0$, or qubits three and four if $x = 1$. The two remaining qubits, labelled $B_1$ and $B_2$, are subsequently transformed by a unitary $V$ that maps $\ket{00}$ to $\ket{0\tau_0}$ and $\ket{11}$ to $\ket{1\tau_1}$ (this is always possible as $\ip{0\tau_0}{1\tau_1} = 0$). $B_1$ is emitted as output while $B_2$ is retained by the transducer as the causal state for the subsequent time-step. Measurement of $B_1$ in the computational basis yields $y$. Iteration of this procedure replicates correct future input-output statistics. The resulting improved efficiency is highlighted in (b), which depicts the maximum memory required by a quantum transducer $\overline{Q}$ (orange surface) to simulated the actively perturbed coin versus its classical counterpart, the structural complexity $\overline{C}$ (blue surface) for various $p$ and $q$. While the $\varepsilon$-transducer generally requires $1$ bit of memory, the quantum transducer requires less, and becomes increasingly more efficient as $p, q \rightarrow 0.5$. }}
\end{figure}

\jt{\textbf{Example Revisited}.  We illustrate these results for the aforementioned actively perturbed coin. Recall that the \et of this process features two causal states, $s_0$ and $s_1$ (see Fig. \ref{fig:causalstatediagram}). As this transducer is step-wise inefficient, a more efficient quantum transducer exists. Specifically, set the quantum causal states to
\begin{equation}\label{eq:perturbedcoinstates}
\ket{\tau_0} = \sqrt{r}\ket{0} + \sqrt{1-r} \ket{1}, \qquad \ket{\tau_1} = \ket{0},
\end{equation}}
\jt{where  $r = 16pq(1-p)(1-q)$. Note that while these states do not resemble the standard form in Eq. $\eqref{eq:quantumcausalstates}$ they are unitarily equivalent. Given $\{\ket{\tau_i}\}_i$, we can initialize the joint four qubit state $\ket{\tau_i000}$ (with three ancilla), and implement an appropriate 4-qubit unitary $U: \ket{\tau_i000} \rightarrow \ket{s_i}$ for $j = 0,1$, where
\begin{equation}\nonumber
\ket{s_0} = \ket{\phi_p}_{12} \otimes \ket{\phi_q}_{34}, \qquad
\ket{s_1} = \hat{X}_1\hat{X}_2\hat{X}_3\hat{X}_4 \ket{s_0}.
\end{equation}
are of standard form. Here subscripts $1\dots 4$ label each of the four qubits, while $\ket{\phi_p} = \sqrt{1-p}\ket{00} + \sqrt{p}\ket{11}$ and $\ket{\phi_q} = \sqrt{1-q}\ket{00} + \sqrt{q}\ket{11}$, and $\hat{X}_i$ represents the Pauli $X$ operator on the $i^{th}$ qubit. The $\ket{\tau_i}$ representation makes it clear that these states can be encoded within a single qubit. Fig. \ref{fig:coinflip}a then outlines a quantum circuit that generates desired future behaviour.}

\jt{The resulting quantum transducer is clearly more efficient. $|\langle \tau_0 | \tau_1\rangle| = \sqrt{r} > 0$, provided $0 < p, q < 1$. Thus $Q_X < C_X$ for all input processes $\omni{X}$. Furthermore, the quantum structural complexity $\bar{Q} = \sup_X Q_X$ is attained for any input process where $\ket{\tau_0}$ and $\ket{\tau_1}$ occur with equiprobability, such as any $\omni{X} = \omni{X}_u$. The improvement is significant, and can be explicitly evaluated (see Fig. \ref{fig:coinflip} (b)). In particular, $\lim_{ p,q \rightarrow 0.5}  Q_X/C_X = 0$. Thus the quantum transducer, in limiting cases, can use negligible memory compared to its classical counterpart.}

\jt{The intuition is that as we approach this limit, the output of the process becomes progressively more random. Thus future black-box behaviour of a process whose last output is heads becomes increasingly similar to one whose last output is tails. The equivalence class $s_0$ or $s_1$, then contains progressively less information about future outputs. A classical transducer nevertheless must distinguish these two scenarios, and thus exhibits an entropy of $1$ when the scenarios are equally likely. A quantum transducer, however, has the freedom to only \emph{partially} distinguish the two scenarios to the extent in which they affect future statistics. In particular  $\lim_{p,q \rightarrow 0.5} |\langle \tau_0 | \tau_1\rangle | \rightarrow 1$. As the process becomes more random, the quantum transducer saves memory by encoding the two scenarios in progressively less orthogonal quantum states.}

\textbf{Future Directions.}\jt{ There is potential in viewing predictive modelling as a communication task, where Alice sends information about a process's past to Bob, so that he may generate correct future statistical behaviour \cite{barnett2015computational,mahoney2016occam}. The simpler a model, the less information Alice needs to communicate to Bob. The entropic benefits of harnessing non-orthogonal causal states mirrors similar advantages in exploiting non-orthogonal codewords to perform certain communication tasks \cite{perry2015communication}. Quantum transducers could thus identify a larger class of such tasks, and provide a general strategy to supersede classical limits. Meanwhile, one may also consider generalisations of statistical complexity that use other measures of entropy. The max entropy, for example, captures the minimum dimensionality required to simulate general input-output processes. This may complement existing work in quantum dimensionality testing~\cite{ahrens2012experimental,PhysRevLett.105.230501,kleinmann2011memory}, pointing to testing the dimensionality of systems by seeing how they transform stochastic processes.}

\jt{Another interesting question is how quantum transducers relate to quantum advantage in  randomness processing \cite{dale2015provable}. In this context, it was shown that quantum sources of randomness (named quoins), in the form $\ket{p} = \sqrt{p} \ket{0} + \sqrt{1-p} \ket{1}$ can be a much more powerful resource for sampling a coin with a $p$-dependent bias $f(p)$, than classical coins of bias $p$. Subsequent experimental implementations have used quoins to sample certain $f(p)$, which are impossible to synthesize when  equipped with only p-coins~\cite{PhysRevLett.117.010502}. Quantum transducers appear to utilize similar effects. The quantum causal states in Eq. \eqref{eq:quantumcausalstates}  also resemble a quantum superposition of classical measurement outcomes, which can be used to generate desired future output statistics more efficiently than classically possible. Is this resemblance merely superficial? Quantum transducers and the quantum Bernoulli factory certainly also field significant differences - both in how they quantify efficiency and in what they consider as input and output. As such this question remains very much open.}

\jt{There could also be considerable interest in establishing what resources underpin the performance advantage in quantum transducers. Non-orthogonal quantum causal states, and thus coherence, is clearly necessary. This non-orthogonality then immediately implies that quantum correlations (in the sense of discord \cite{10.1103/RevModPhys.84.1655}) necessarily exist between the state of the transducer and its past outputs. Could the amount of such resources be related quantitatively to the quantum advantage of a particular transducer? Whether more stringent quantum resources, such as entanglement, are also required at some point to generate correct future statistics, also remains an open question. Certainly, all quantum transducers described here exploit highly entangling operations to generate statistically correct future behaviour - and field a significant amount of entanglement during their operation. Is the existence of such entanglement at some stage during the simulation process essential for quantum advantage?}

\section{Discussion} In computational mechanics, $\varepsilon$-transducers are the provably simplest models of input-output processes. Their internal entropy is a quantifier of structure -- any device capable of replicating the process's behaviour must track at least this much information. Here, we generalize this formalism to the quantum regime. We propose a systematic method to construct quantum transducers  that are generally simpler than their simplest classical counterparts; in the sense that quantum transducers store less information whenever it is physically possible to do so. Our work indicates the perceived complexity of input-output processes generally depends on what type of information theory we use to describe them.

\jt{A natural continuation is to explore the feasibility of such quantum transducers in real world conditions. A proof of principle demonstration is well within reach of present-day technology. The quantum transducer for the actively perturbed coin can be implemented by a single qubit undergoing one of two different weak measurements at each time-step. To demonstrate a quantum advantage for real world applications would also motivate new theory. For example, noise will certainly degrade the performance of quantum transducers, forcing the use of more distinguishable quantum causal states. The derivation of bounds on the resultant entropic cost would thus help us establish thresholds that guarantee quantum advantage in real-world conditions.}

Ultimately, the interface between quantum and computational mechanics motivates the potential for tools of each community to impact the other. Classical transducers, for example, are used to capture emergence of complexity under evolutionary pressures~\cite{crutchfield2006objects}, distil structure within cellular automata~\cite{hanson1992attractor,shalizi2001causal}, and characterize the dynamics of sequential learning~\cite{crutchfield2012structural}, and it would be interesting to see how these ideas change in the quantum regime. Meanwhile, recent results suggests the inefficiency of classical models may incur unavoidable thermodynamic costs \cite{wiesner2012information,garner2015simpler,cabello2015thermodynamical}. In reducing this inefficiency, quantum transducers could offer more energetically efficient methods of transforming information. Any such developments could demonstrate the impact of quantum technologies in domains where their use has not been considered before.

\section{Methods}

\textbf{Definitions.} Let $X^{(t)}$,$Y^{(t)}$ and $S^{(t)}$ represent respectively the random variables governing the input, output and causal state at time $t$. Let $Y^{(0:t)} = Y^{(0)}\dots Y^{(t)}$ govern the outputs $y^{(0:t)} = y^{(0)}\dots y^{(t)}$  and $X^{(0:t)} = X^{(0)}\dots X^{(t)}$ govern the inputs $x^{(0:t)} = x^{(0)}\dots x^{(t)}$. We introduce ordered pairs $Z^{(t)} = (X^{(t)}, Y^{(t)})$ which take values $z^{(t)} = (x^{(t)}, y^{(t)}) \in \mathbfcal{Z}$, where $\mathbfcal{Z}$ represents the space of potential input-output pairs. In analogy, let $\past{z} = (\past{x}, \past{y})$ represents a particular past, $\past{\mathbfcal{Z}}$ be the space of all possible pasts and $z^{(0:t)} = z^{(0)}\dots z^{(t)} \in \mathbfcal{Z}^{t+1}$ the input-output pairs from time-steps $0$ to $t$. Let $|\mathbfcal{X}|$ denote the size of the input alphabet, $|\mathbfcal{Y}|$ the size of output alphabet and $n = |\mathbfcal{S}|$ the number of causal states \jt{(if finite)}. Without loss of generality, we assume the inputs and outputs are labeled numerically, such that $\mathbfcal{X} = \{x_i\}_0^{|\mathbfcal{X}|-1}$ and $\mathbfcal{Y} = \{y_i\}_0^{|\mathbfcal{Y}|-1}$.

Each instance of an input-output process $\{\omni{Y}|\omni{x}\}$ exhibits a specific past $\past{z} = (\past{x},\past{y})$. When supplied $x^{(0)} = x$ as input, it emits $y^{(0)} = y$ with probability $P[Y^{(0)}=y| X^{(0)} = x, \past{Z} = \past{z}]$. The system's past then transitions from $\past{z}$ to $\past{z}' = (\past{x}x, \past{y}y) = (\past{x}', \past{y}')$. This motivates the \emph{propagating functions} $\mu_{x,y}$ on the space of pasts, $\mu_{x,y}(\past{z}) = (\past{x}x, \past{y}y)$, which characterize how the past updates upon observation of $(x,y)$ at each time-step. Iterating through this process for $t+1$ timesteps gives expected output $y^{(0:t)}$ with probability $P(Y^{(0:t)} = y^{(0:t)}| X^{(0:t)}=x^{(0:t)}, \past{Z} = \past{z})$ upon receipt of input sequence $x^{(0:t)}$. Taking $t \rightarrow \infty$ gives the expected future input-output behaviour $P[\future{Y} | \future{X} = \future{x}, \past{Z} = \past{z}]$ upon future input sequence $\future{x} \in \future{\mathbfcal{X}}$.

A quantum model of an input-output process defines an encoding function $\aleph$ that maps each past $\past{z}$ to some state $\aleph(\past{z}) = \rho_{\past{z}}$ within some physical system $\Xi$.  The model is correct, if there exists a family of operations $\mathbb{M} = \{\mathbfcal{M}_{x}\}_{x \in \mathbfcal{X}}$ such that application of $\mathbfcal{M}_{x}$ on $\Xi$ replicates the behaviour of inputting $x$. That is, $\mathbfcal{M}_x$ acting on $\rho_{\past{z}}$ should (1) generate output $y$ with probability $P[Y^{(0)} | X^{(0)} =x, \past{Z} = \past{z}]$ and (2) transition $\Xi$ into state $\aleph[\mu_{x,y}(\past{z})] = \rho_{\mu_{x,y}(\past{z})}$. (1) ensures the model outputs statistically correct $y^{(0)}$, while (2) ensures the model's internal memory is updated to record the event $z^{(0)} = (x^{(0)}, y^{(0)})$, allowing correct prediction upon receipt of future inputs. Sequential application of $\mathbfcal{M}_{\future{x}} = \mathbfcal{M}_{x^{(0)}},\mathbfcal{M}_{x^{(1)}},\ldots$ then generates output $\future{y}$ with probability $P[\future{Y}| \future{X} = \future{x}, \past{Z} = \past{z}]$. Let $\Omega = \{\rho_{\past{z}}\}_{\past{z} \in \past{\mathbfcal{Z}}}$ be the image of $\aleph$. We now define quantum models as follows.

\jt{\begin{definition}A general quantum model of an input-output process is a triple $\mathbfcal{Q} = (\aleph, \Omega, \mathbb{M})$, where $\aleph$, $\Omega$ and $\mathbb{M}$ satisfy the conditions above.
\end{definition}}

Each stationary input process $\omni{X}$ induces a probability distribution $p_{X}(\past{z})$ over the set of pasts, and thus a steady state of the machine $\rho_X$. The resulting entropy of $\Xi$,
\begin{equation}
Q_X[\mathbfcal{Q}] = -{\rm Tr}\left({\rho_X \log \rho_X}\right),
\end{equation}
then defines the model's input-dependent complexity. For the quantum transducers in the main body, \jt{$\Omega = \{\ket{s_i}\bra{s_i}\}_i$} corresponds to the set of quantum causal states, and the encoding function $\aleph$ maps each past $\past{z}$ to \jt{$\ket{s_i}\bra{s_i}$}, whenever $\past{z} \in s_i$. Specifically, if we define the classical encoding function $\epsilon: \past{\mathbfcal{Z}} \rightarrow \mathbfcal{S}$ that maps pasts onto causal states such that $\epsilon(\past{z}) = s_i $ iff $\past{z} \in s_i$, then $\ket{s_i}  = \ket{\epsilon(\past{z})}$.

We also introduce \emph{input strategies}. Suppose Bob receives a model of some known input-output process, initialized in some (possibly unknown) state $\rho_{\past{z}} \in \Omega$. Bob now wants to drive the model to exhibit some particular future behaviour by using an input strategy -- a specific algorithm for deciding what input $x^{(t)}$ he will feed the model at each specific $t \geq 0$, purely from the model's black-box behaviour.

\begin{definition}An input strategy is a family of functions $F = \{f^{(t)} | t \in \mathbb{Z}^+\}$, where $f^{(t)}: \mathbfcal{Z}^{t} \rightarrow \mathbfcal{X}$ is a map from the space of pre-existing inputs-outputs onto the input at time $t$, such that $x^{(t)} = f^{(t)}(z^{(0:t-1)})$. We denote a sequence of future inputs which is determined using $x^{(t)} = f^{(t)}(z^{(0:t-1)})$, as $\future{x}_F$.\end{definition}

In subsequent proofs, we will invoke input strategies on classical $\varepsilon$-transducers. Here, we denote
\begin{equation}
\begin{gathered}
P_{s_i, F}[ \future{y}]  =  P [\future{Y} = \future{y}|, \future{X} = \future{x}_F,  S^{(-1)} =s_i],
\end{gathered}
\end{equation}
as the probability distribution that governs future outputs $\future{Y}$, when an input strategy $F$ is used to select the future inputs $\future{x}_F$ to an $\varepsilon$-transducer initialized in some causal state $s_i \in \mathbfcal{S}$.

We also make use of the \emph{trace distance}, $D[P,Q] = \frac{1}{2}\sum_{\future{y}\in\future{\mathbfcal{Y}}} |P[\future{y}] - Q[\future{y}]|$ between two probability distributions $P[\future{y}]$ and $Q[\future{y}]$. Similarly any two quantum states $\tau$ and $\sigma$ have trace distance $D[\tau, \sigma] = \frac{1}{2}\rm{Tr}|\tau - \sigma|$, where $|A| \equiv \sqrt{A^{\dagger} A}$.

{\bf Proof of Correctness.} Here, we prove a quantum transducer can generate correct future statistical behaviour when supplied with a quantum system $\Xi$ initialized in state \jt{$\ket{s_i}$}, encoding information about $\past{z} \in s_i$. That is, there exists a family of quantum processes $\mathbb{M} = \{\mathbfcal{M}_{x}\}$ whose action on $\Xi$ produces an output $y$ sampled from  $P[Y^{(0)} \! = \! y | X^{(0)} = x, S^{(-1)} \! =\!  s_i] $, while transforming the state of $\Xi$ to \jt{$ \ket{\varepsilon(\past{x} x, \past{y} y)} $.}

\emph{Proof.} Recall that $\ket{s_i} = \bigotimes_{x} \ket{s_i^x}$ where $\ket{s_i^x} = \sum_{k =0}^{n-1}  \sum_{y \in \mathbfcal{Y}} \sqrt{T^{y|x}_{ik} }\ket{y}\ket{k}$. Let $\Sigma$ be the $|\mathbfcal{Y}|$ dimensional Hilbert space spanned by $\{\ket{y}\}_{y\in \mathbfcal{Y}}$ and $\mathcal{K}$ be the $n$-dimensional Hilbert space spanned by $\{\ket{k}\}_{k=0}^{n-1}$. Then each $\ket{s_i^x}$ lies in $\omega = \Sigma \otimes \mathcal{K}$, and each causal state $\ket{s_i}$ lies within $\mathcal{W}= \omega^{\otimes |\mathbfcal{X}|}$. The set  $\{\ket{s_i}\}_{i=0}^{n-1}$ spans some subspace of $\mathcal{W}$ of dimension at most $n$. This implies that the causal states can be stored within $\mathcal{K}$ without loss, i.e., there exists quantum states $\{\ket{\tau_i}\}_{i = 0}^{n-1}$ in $\mathcal{K}$ and a unitary process $U$ such that $U : \ket{\tau_i} \rightarrow  \ket{s_i}$ for all $i$ (Note, this assumes appending suitable ancillary subspaces to each $\ket{\tau_i}$). An explicit form for $\ket{\tau_i}$ can be systematically constructed through Gram-Schmidt decomposition~\cite{noble1988applied}. We refer to $\ket{\tau_i}$ as \emph{compressed causal states}, and $U$ as the \emph{decompression operator}.

\jt{We define the \emph{selection operator} $S_x: \mathcal{W} \rightarrow \omega$ such that \jt{$S_{x'}:\, \left(\bigotimes_{x} \ket{\phi_{x}}\right) \rightarrow \ket{\phi_{x'}}$.}  Physically, if $\mathcal{W}$ represents a state space of $|\mathbfcal{X}|$ qudits each with state space $\omega$ labelled from $0$ to $|\mathbfcal{X}|-1$, $S_x$ represents discarding \jt{(or tracing out)} all except the $x^{th}$ qudit. Meanwhile, let $B$ be a quantum operation on $\omega$ such that $B: \ket{y}\ket{k}\bra{y}\bra{k} \rightarrow \ket{y}\ket{\tau_k}\bra{y}\bra{\tau_k}$. This operation always exists as it can be implemented by Kraus operators $E_{yk}= \ket{y}\ket{\tau_k} \bra{y}\bra{k}$.}

\begin{figure}
\includegraphics[width=0.5\textwidth]{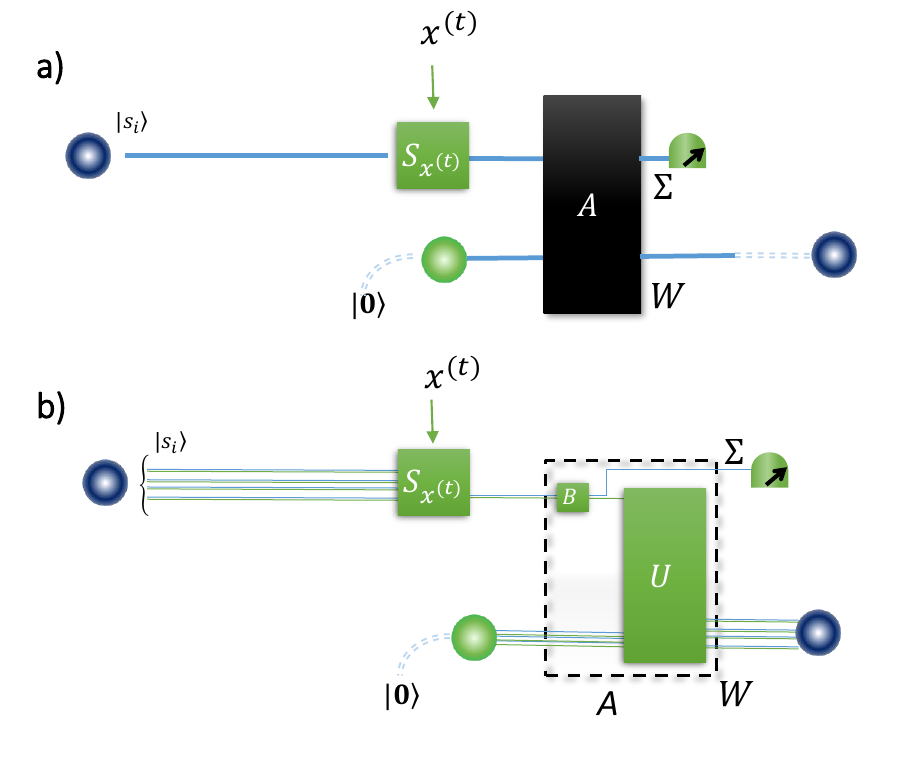}
\caption{\label{fig:circuitdetails} \jt{(a) The process by which the quantum transducer generates future statistics as outlined in Fig. \ref{fig:quantumsimulation}. (b) A more detailed breakdown of Fig. \ref{fig:circuitdetails} (a). Upon input $x$ at time $t$, the transducer first applies the selection operator $S_x$.  The subsequent operator $A$ can be decomposed into two operators, a linear mapping $B : \ket{s_i^x}\bra{s_i^x} \rightarrow \sum_{y,k} T^{y|x}_{ik} \ket{y} \ket{\tau_k} \bra{y}\bra{\tau_k}$, and a decompression operator $U$ that rotates each  $\ket{\tau_k}$  into $\ket{s_k}$ (always possible when suitable ancillary systems in states ${\bf \ket{0}}$ are supplied). $\Sigma$ is emitted as output, while $\mathbfcal{W}$ is retained as the subsequent causal state at time $t$. This circuit makes it clear that $\ket{\tau_k}$   also make perfectly valid causal states.}}
\end{figure}
\jt{The quantum transducer operates as follows. Upon input $x$ at time $t$, it applies $S_x$ on $\Xi$, followed by execution of $B$. This transforms the state of the system from $\ket{s_i^{x}}$ to $\sum_{k=0}^{n-1} \sum_{y \in \mathbfcal{Y}} T^{y|x}_{ik} \ket{y} \ket{\tau_k} \bra{y} \bra{\tau_k} $. Application of the decompression operator then gives $\sum_{k=0}^{n-1} \sum_{y \in \mathbfcal{Y}} T^{y|x}_{ik} \ket{y} \ket{s_k} \bra{y} \bra{s_k}$ on state space $\Sigma \otimes \mathcal{W}$. The machine then emits $\Sigma$ as the output. Measurement of $\Sigma$ by an external observer in basis $\{\ket{y}\}_{y \in \mathbfcal{Y}}$ gives the output $y$ at time $t$, while inducing $\mathcal{W}$ to transition to the subsequent quantum causal state. $\mathcal{W}$ is then retained in $\Xi$.}

The above procedure establishes a family of quantum operations $\{\mathbfcal{M}_x\}_{x \in \mathbfcal{X}}$ that maps each quantum causal state $\ket{s_i}$ to $\ket{s_j}$ while emitting output $y$ with probability $T_{ij}^{y|x}$. Thus the quantum transducer operates statistically indistinguishably from its classical counterpart, and must be correct. To simulate the future behaviour of the input-output process upon future input $\future{x}$, the quantum transducer iterates through the above process by sequential application of $\mathbfcal{M}_{x^{(0)}}$, $\mathbfcal{M}_{x^{(1)}}$, etc. This establishes the transducer is a valid quantum model.

\textbf{Proof of Reduced Complexity and Generality.} Given an arbitrary input-output process, whose $\varepsilon$-transducer and quantum transducer has respective input dependent complexities of $C_X$ and $Q_X$, we show that
\begin{itemize}
\item{\emph{Reduced Complexity:} $Q_X  < C_X$ for all non-pathological input processes $\omni{X}$, whenever the $\varepsilon$-transducer is step-wise inefficient.}
\item \emph{Generality:} Given an input-output process, either $Q_X < C_X$ for all non-pathological input processes $\omni{X}$ or there exists no physically realizable model that does this.
\end{itemize}
Our strategy makes use of the following theorem.
\begin{theorem}
For any input-output process the following statements are equivalent:
\begin{center}
 \begin{tabular}{l}
(I) $\langle s_i | s_j\rangle = 0$ for any $\ket{s_i}, \ket{s_j}\in \{\ket{s_i}\}_i$ where $i\neq j$. \\
(II) For any pair $(s_i, s_j) \in \mathbfcal{S} \times \mathbfcal{S}$ where $i\neq j$, $\exists x \in \mathbfcal{ X}$ \\ such that $\forall y \in \mathbfcal{Y}, s_k \in \mathbfcal{S}$, the product $T^{y|x}_{ik} T^{y|x}_{jk} = 0$.\\
(III) For any pair $(s_i, s_j) \in \mathbfcal{S} \times \mathbfcal{S}$ where $i\neq j$,  there \notag \\ exists  an  input strategy $F$  such that $D[P_{s_i, F}, P_{s_j, F}] = 1$.\\
(IV) For any input process $\omni{X}$, any physically realizable \\ (quantum) model must have complexity at least $C_X$.
 \end{tabular}
\end{center}
\end{theorem}
 To prove this theorem, we show {\bf A.} (I) is equivalent to (II). {\bf B.} (II) implies (III), {\bf C.} (III) implies (IV) and {\bf D.} (IV) implies (I).

\emph{Proof of \textbf{A.}} We prove this by showing (I) is false iff (II) is false. First, assume (I) is false, such that there exists $\ket{s_i}, \ket{s_j} \in \{\ket{s_i}\}_i$ with $ \langle s_i | s_j \rangle = \Pi_{x \in \mathbfcal{X}} \left(\sum_{y, k} \sqrt{T^{y | x}_{i k} T^{y |x}_{jk}}\right) > 0 $. This implies that for all $x \in \mathbfcal{X}$ there exists $s_k \in \mathbfcal{S}, y \in \mathbfcal{Y}$ such that $T^{y |x}_{i k} T^{y |x}_{j k} \neq 0$. Thus (II) is false.  Meanwhile assume (II) is false, i.e., there exists $s_i$ and $s_j$  such that $\forall x$ we can find $ y_x \in \mathbfcal{Y}, s_{k_x} \in \mathbfcal{S}$ for which $T^{y_x|x}_{ik_x} T^{y_x|x}_{jk_x} > 0$. It follows that $\langle s_i | s_j \rangle > \Pi_{x} \left(T^{y_x|x}_{ik_x} T^{y_x|x}_{jk_x}\right) > 0$ and (I) is false.

\emph{Proof of \textbf{B.}} To prove this, we introduce the update function $g: \mathbfcal{S}\times\mathbfcal{X}\times \mathbfcal{Y} \rightarrow \mathbfcal{S}$, such that $g(s_i, x, y) = s_k$ iff $T^{y|x}_{ik} \neq 0$. Note that $g$ is always a function by joint unifiliarity of the $\varepsilon$-transducer~\cite{barnett2015computational}. That is, the triple $(s^{(t-1)},x^{(t)}, y^{(t)})$ uniquely determines $s^{(t)}$.

\jt{We also use the following game to elucidate the proof.} At time $t= -1$, Alice initializes an $\varepsilon$-transducer in either $s_i$  or $s_j$ and seals it inside a black box. Alice gives this box to Bob and challenges him to infer whether $S^{(-1)} = s_i$ or $S^{(-1)} = s_j$, based purely on the transducer\jt{'}s future black-box behaviour. We first prove that if (II) is true, then for each pair $(s_i,s_j)$ there exists an input strategy $F_{ij}$ that allows Bob to discriminate $S^{(-1)} = s_i$ from $S^{(-1)} = s_j$ to arbitrary accuracy purely from the transducer's output behaviour.

Specifically (II) implies that for all pairs $(s_i, s_j) \in \mathbfcal{S} \times \mathbfcal{S}$,  there exists some $x_{ij}$ such that $\forall y \in \mathbfcal{Y}, s_k \in \mathbfcal{S}$, $T^{y|x_{ij}}_{ik} T^{y|x_{ij}}_{jk} = 0$. At $t=0$, Bob inputs $X^{(0)} = x_{ij}$. Let $y = Y^{(0)}$ be the corresponding output. Note that because we have observed $y = Y^{(0)}$, there must have been some non-zero probability of observing $y$ on input $x_{ij}$, i.e., at least one of (i) $P[ Y^{(0)} \! = \! y|  X^{(0)} \! = \! x_{ij}, S^{-1} \! =\!  s_i] \neq 0$ or (ii) $P[ Y^{(0)} \! =\!  y|   X^{(0)} \! =\!  x_{ij}, S^{-1} \! =\!  s_j] \neq 0$ must be true.  This presents two different possible scenarios:
\begin{itemize}
\item[(a)]{Only one of (i) and (ii) is true. That is, one of $s_i$ or $s_j$ never outputs $y$ upon input $x_{ij}$ -- such that either $P[ Y^{(0)} \! = \! y|  X^{(0)} \! = \! x_{ij}, S^{-1} \! =\!  s_i] = 0$ or $P[ Y^{(0)} \! =\!  y|   X^{(0)} \! =\!  x_{ij}, S^{-1} \! =\!  s_j]  = 0$.}
\item[(b)]{Both (i) and (ii) are true, implying $g(s_i,y,x_{ij}) \neq g(s_j,y,x_{ij})$.}
\end{itemize}
If (a) occurs, then Bob can immediately determine whether $S^{(-1)} = s_i$ or $S^{(-1)} = s_j$ and we are done. If (b) occurs, let $s_i' = g(s_i, y, x_{ij})$ be the new causal state if $S^{(-1)} = s_i$. Let $s_j' = g(s_j, y, x_{ij})$ be the new causal state if $S^{(-1)} =s_j$. Due to joint unifiliarity of the $\epsilon$-transducer, Bob is able to uniquely determine $s_j'$ and $s_i'$ upon observation of $z^{(0)} = (x_{ij}, y)$. Given $s_i'$ and $s_j'$, (II) implies Bob can find $x_{ij}'$ such that $\forall y \in \mathbfcal{Y}, s_k \in \mathbfcal{S}$ the product $T^{ y|x_{ij}'}_{i'k} T^{y|x_{ij}'}_{j'k} = 0$. Thus we can repeat the steps above choosing $x^{(1)} = x_{ij}'$. Iterating this procedure defines an input strategy $F_{ij}$, which determines each input $x^{(t)}$  as a function of observed inputs and outputs. At each point in time $t$, Bob will be able to identify some $s^{(t)}_i$ which is the current causal state if $S^{(-1)} = s_i$, and some $s^{(t)}_j$ which is the current causal state if $S^{(-1)} = s_j$.

Eventually, either scenario (a) will occur allowing Bob to perfectly rule out $S^{(-1)} = s_i$. Alternatively in the limit of an infinite number of time steps, Bob can synchronize the $\varepsilon$-transducer based on the observed inputs and outputs (that is, the causal state at time $t$ is entirely determined by observation of the past in limit of large $t$~\cite{shalizi2001computational,barnett2015computational}). Thus Bob can determine $s^{(t)}$ in the limit as $t \rightarrow \infty$, allowing inference of whether $S^{(-1)} =s_i$ or $S^{(-1)} =s_j$. This constitutes an explicit input strategy $F_{ij}$ that allows Bob to discriminate between $S^{(-1)} =s_i$ and $S^{(-1)} = s_j$ to any arbitrary accuracy. Accordingly $D[P_{s_i F_{ij}}, P_{s_j, F_{ij}}] = 1$.

\emph{Proof of \textbf{C.}} We prove this via its contrapositive. That is, suppose (IV) is false, such that there exists a quantum model $\mathbfcal{Q}'$ with identical input-output relations to the process's $\varepsilon$-transducer, which stores $Q_X[\mathbfcal{Q}'] < C_X$ for some $\omni{X}$. We show that if (III) is true then the data processing inequality is violated~\cite{nielsen2010quantum}.

We first make use of the following observation: If a model $\mathbfcal{Q}' = (\aleph', \Omega', \mathbb{M}')$ satisfies $\aleph' (\past{z}) \neq \aleph' (\past{z}')$ for some $ \past{z} \sim_{\varepsilon} \past{z}' $, then we can always construct an alternative model $\mathbfcal{Q} = (\aleph, \Omega, \mathbb{M})$ such that $Q_X[\mathbfcal{Q}] \le Q_X [\mathbfcal{Q}']$ for all input processes $\omni{X}$, and $\aleph(\past{z}) = \aleph(\past{z}')$ iff $\varepsilon(\past{z}) = \varepsilon(\past{z}')$ for all $\past{z}, \past{z}' \in \past{\mathbfcal{Z}}$. (This is a consequence of the concavity of entropy, see methods in \cite{suen2015classical}). I.e., for any model $\mathbfcal{Q}'$ with quantum states $\Omega'$ not in 1-1 correspondence with classical causal states, there always exists a simpler model $\mathbfcal{Q}$ whose quantum states are in 1-1 correspondence with the causal states.

Thus, falsehood of (IV) implies there must exist some quantum model $\mathbfcal{Q} = (\aleph, \Omega, \mathbb{M})$ such that (i) $\aleph(\past{z}) = \aleph(\past{z}')$ if and only if $\past{z} \sim_{\varepsilon} \past{z}'$ and (ii) $Q_X[\mathbfcal{Q}] < C_X$ for some $\omni{X}$. Now by virtue of (ii), there must exist two states $\rho_i, \rho_j \in \Omega$ such that the trace distance
\begin{equation}
D[\rho_i,\rho_j] < 1.
\end{equation}
The data processing inequality therefore implies that any quantum operation $\mathbfcal{M}_{\future{x}}: \Xi \rightarrow \future{\mathbfcal{Y}}$ that generates future output statistics must satisfy $D[\mathbfcal{M}_{\future{x}}(\rho_i), \mathbfcal{M}_{\future{x}}(\rho_j)] \le D[\rho_i, \rho_j] < 1.$

However,  all models of the same input-output process have identical black-box behaviour. In particular the $\varepsilon$-transducer of the input-output process that $\mathbfcal{Q}$  models, must behave identically to $\mathbfcal{Q}$. As such, there exists two causal states of the classical $\varepsilon$-transducer, $s_i, s_j \in \mathbfcal{S}$ such that $D[P_{s_i, F}, P_{s_j, F}] = D[\mathbfcal{M}_{\future{x}_F} (\rho_i), \mathbfcal{M}_{\future{x}_F} (\rho_j)]  < 1$  for all possible input strategies $F$. This implies (III) is false. Thus we have used proof by contrapositive to show (III) implies (IV).

\emph{Proof of \textbf{D.}} The quantum transducer is a physically realizable model. Thus (IV) implies that $Q_X \geq C_X$ for all $\omni{X}$. However, we note from our construction $Q_X \leq C_X$ for all $\omni{X}$. Therefore $Q_X = C_X$. Since the causal states of the quantum transducer are all pure, all $\ket{s_i}$ are mutually orthogonal~\cite{nielsen2010quantum}.

\emph{Proof of Main Result.} Reduced complexity and generality are consequences of the above theorem. Specifically given a particular input-output process, falsehood of (II) implies that its transducer is step-wise inefficient. Meanwhile, falsehood of (I) implies $Q_X < C_X$ for all non-pathological $\omni{X}$. Thus reduced complexity is implied by equivalence of (I) and (II). Generality is proven by contradiction. Assume that for some non-pathological $\omni{X}$, quantum transducers yield no improvement (i.e. $Q_X = C_X$) but some other physically realizable model has complexity less than $C_X$. The former implies (I) is true, the latter implies (IV) is false, which violates the theorem. Thus, both reduced complexity and generality must hold.

\textbf{Acknowledgements.} We thank Suen Whei Yeap, Blake Pollard, Liu Qing, and Yang Chengran for their input and helpful discussions.

\textbf{Conflict of Interest Declaration.} There are no conflicts of interest involved in the production of this manuscript.

\textbf{Contributions.} All authors conceptualized the project and developed the examples.  J.T. worked through the detailed calculations. J.T. and M.G wrote the manuscript. M.G. led the project.

\textbf{Funding.} This work was funded by the John Templeton Foundation Grant 53914 {\em ``Occam's Quantum Mechanical Razor: Can Quantum theory admit the Simplest Understanding of Reality?''}; the Oxford Martin School;
 the Ministry of Education in Singapore, the Academic Research Fund Tier 3 MOE2012-T3-1-009; the Foundational Questions Institute Grant Observer-dependent complexity: The quantum-classical divergence over ‘what is complex?’; the National Research Foundation of Singapore and in particular NRF Award No. NRF--NRFF2016--02.

\end{document}